\documentstyle[aps]{revtex}
\begin{document}
\begin{flushright}
Preprint IHEP 96-\\
August 1996\\
hep-ph/9608435
\end{flushright}
\begin{center}
{\large\bf Decay of $B_c^{*+}(3S)\to B^+D^0$}\\
\vspace*{3mm}
V.V.~Kiselev\\
Institute for High Energy Physics,\\
Protvino, Moscow Region, 142284, Russia\\
E-mail: kiselev@mx.ihep.su~~~~~~  Fax: +7-095-2302337
\end{center}
\abstract
{The decay constant for the vector state of $3S$-level in the heavy
$(\bar b c)$-quarkonium is evaluated in the framework of sum rules for
the mesonic currents. A scaling relation for the constants of vector 
quarkonia with different quark contents is derived. The numerical estime 
gives $\Gamma(B_c^{*+}(3S)\to B^+D^0)=90 \pm 35$ MeV.}
\vspace*{3mm}

PACS numbers: 13.25.Gv
\section{Introduction}
The experimental search for the $B_c^+$ meson in the facilities with the
vertex detectors (OPAL \cite{1}, ALEPH \cite{2}, DELPHI\cite{3} and CDF
\cite{4}) stimulated the theoretical studies on the spectroscopy of the
heavy $(\bar bc)$-quarkonium \cite{5}, mechanisms of its production in 
different interactions \cite{6} and on estimates of different decay widths
for the both basic state \cite{7} and excited levels \cite{5,8}. The feature
of the $(\bar bc)$-system is the absense of the annihilation decay modes
caused by the strong or electromagnetic interactions. 
So, the basic pseudoscalar
$B_c^+$ state decays due to the weak interaction, and it is the long-lived
particle, $\tau(B_c^+)= 0.55\pm 0.15$ ps \cite{7,9}. The excited 
$(\bar b c)$-quarkonium levels lying below the threshold of the decay to the
heavy meson $BD$ pair, radiatively transform into the $(\bar b c)$-states
with the smaller masses. The $B_c^{*+}(3S)$ state is above the $BD$
thershold, so its decay is analogous to $\Upsilon(4S)\to B^+B^-$.
The constant of the latter decay was considered in ref.\cite{10} in the 
framework of the sum rules for the mesonic currents.

In this work  we consider the $g$ constant for the decay of the vector
quarkonium, generally containing the quarks of different flavors,
say, $(\bar b c)$ for the definite notations. This heavy quarkonium with the
mass $M$, satisfying the condition $m_B+m_D < M < m_{B^*}+m_{D^*}$, decays to
the heavy meson pair $B^+D^0$. We derive the scaling relation
$$
\frac{g^2}{M}\; \biggl(\frac{4\mu_{BD}}{M}\biggr) = const.\;,
$$
where $\mu_{BD}= m_B m_D/(m_B+m_D)$ is the reduced mass of the heavy meson 
pair. The constant value in the right hand side of the relation is 
the same for the decays of $\Upsilon(4S)\to B^+B^-$, $B_c^{*+}(3S)\to B^+D^0$
and $\psi(3770)\to D^+D^-$, where $\mu_{BB}=M_{\Upsilon(4S)}/4$,
$\mu_{DD}=M_{\psi(3770)}/4$.

In Section II we consider the sum rules for the mesonic currents. 
In Section III the scaling relation is derived and numerical estimates are
performed. In the Conclusion the obtained results are summarized.

\section{Sum rules}
Let us consider the vector current of mesons
$$
J_\mu^{BD}(x) = \frac{i}{2}[B^+(x)\cdot \partial_\mu D^0(x) -
\partial_\mu B^+(x)\cdot D^0(x)]
$$
and define the contribution of this current into the leptonic $f_{BD}$ 
constant of the vector $(\bar b c)$-quarkonium lying above the $BD$-threshold
\begin{equation}
i f_{BD} M \epsilon^{(\lambda)}_\mu \; e^{ipx} =
\langle 0|J_\mu^{\dag BD}(x)|V_{(\bar b c)},\lambda\rangle\;,
\label{f}
\end{equation}
where $\lambda$ is the polarization of the $V_{(\bar b c)}$ state,
$\epsilon_\mu^{(\lambda)}$ is its vector of polarization, $p$ is the
$V_{(\bar b c)}$ momentum, $p^2=M^2$.

Further, introduce the $\cal F$ form factor for the 
transversal interaction of the $BD$ pair with the vector ${\cal A}_\mu$ 
current due to the vertex
\begin{equation}
{\cal L}^{tr}_{J\cal A} = {\cal F}(q^2)\; {\cal A}_\mu\cdot k^\mu\;,
\label{Lj}
\end{equation}
where $q=p_B+p_D$, $p_{B,D}$ are the momenta of the meson lines directed
out the vertex, and $p_B=q_B+k$, $p_D=q_D-k$, $q_{B,D}\cdot k=0$.
Thus, one has
$$
\biggl(g^{\mu\nu}-\frac{q^\mu q^\nu}{q^2}\biggr)
\langle 0|J^{\dag BD}_\nu(0)|B^+(p_B)D^0(p_D)\rangle = i\; 
{\cal F}(q^2)\; k^\mu\;.
$$

Consider the transversal part of the current correlator, 
$$
\Pi^{tr}_{JJ}(q^2) =
\frac{1}{3}\biggl(-g^{\mu\nu}+\frac{q^\mu q^\nu}{q^2}\biggr)
\int d^4x e^{iqx}\langle 0|T\; J_\mu^{\dag BD}(x) J_\nu^{BD}(0)|0\rangle\;.
$$
One can isolate the contribution of the resonance lying above the kinematical
threshold of the $BD$ pair, so that
$$
\Pi^{tr}_{JJ}(q^2) = \frac{f^2_{BD} M^2}{M^2-q^2} +
\int^{\infty}_{s_{th}} \frac{ds}{s-q^2} \rho(s)\;,
$$
where $\rho(s)$ is the density of the nonresonant contribution. 
On the other hand, the form factor in (\ref{Lj}) determines the value
\begin{equation}
\Im m \Pi^{tr}_{\cal FF}(q^2) = 
\frac{1}{8\pi}\; \frac{|{\bf k}|^3}{3\sqrt{q^2}}\; {\cal F}^2(q^2)\;,
\label{FF}
\end{equation}
where $|{\bf k}|^2=-k^2=(q^2+m_B^2-m_D^2)^2/(4q^2)-m_B^2$.
Write down the sum rules for the mesonic currents
$$
\Pi^{tr}_{JJ}(q^2) = \frac{1}{\pi} \int^{\infty}_{s_i}
\frac{ds}{s-q^2} \Im m \Pi^{tr}_{\cal FF}(s)\;,
$$
where $s_i=(m_B+m_D)^2$. One can consider the following model for the 
continuum density in the form
$$
\rho(s)= \frac{1}{\pi} \Im m \Pi^{tr}_{\cal FF}(s)\; \theta(s-s_{th})\;.
$$
Then the sum rules are given by the following expression
\begin{equation}
\frac{f^2_{BD} M^2}{M^2-q^2} = \frac{1}{\pi}
\int^{s_{th}}_{s_i} \frac{ds}{s-q^2} \Im m \Pi^{tr}_{\cal FF}(s)\;.
\label{SR}
\end{equation}
The value of the continuum threshold is determined by the energy of 
new channels in the particle production by the $J_\mu$ current. As was shown in
\cite{10} for the $\Upsilon(4S)\to B^+B^-$ and $\psi(3770)\to D^+D^-$
decays, this value is given by the threshold of production
of the vector $B^{*+}B^{*-}$ and $D^{*+}D^{*-}$ states, so that
we suppose
$$
s_{th}=(m_{B^*}+m_{D^*})^2\;.
$$
Define
$$
v^2(s)= 1- \frac{4m_Bm_D}{s-(m_B-m_D)^2}\;.
$$
Then one has $v^2_{th}\ll 1$.

Further, the consideration of the $\cal F$ form factor in a model for the
$B^+B^-$ and $D^+D^-$ currents \cite{10} resulted in the fact that relation 
(\ref{SR}) and its initial four derivatives over $q^2$ at $q^2=0$ give the
stable value of $f$ with the accuiracy of 5\% to 25\%, correspondingly.
Allowing for the mentioned region of applicability (the number of
the spectral density moment is less than 5), one can transform the integration 
in (\ref{SR}) to the variable of $v^2(s)$ and suppose $q^2=0$ and
${\cal F}(s)\approx {\cal F}(s_i)= F$. Then at $v^2_{th}\ll 1$ and
$|{\bf k}|\approx 2\mu_{BD} v$, one has
$$
f^2_{BD}\approx\frac{1}{\pi}\int^{v_{th}}_{0} dv^2 \cdot v^3\;
\biggl(\frac{4\mu_{BD}}{M}\biggr)^4\; \frac{F^2}{64\pi}\; \frac{M^2}{3}\;.
$$
So
\begin{equation}
f_{BD}= \frac{FM}{4\pi} \biggl(\frac{4\mu_{BD}}{M}\biggr)^2
\sqrt{\frac{v^5_{th}}{30}}\;.
\label{F1}
\end{equation}
Introduce the transversal vertex of the $V_{(\bar b c)}$ state decay to 
the $B^+D^0$ pair
\begin{equation}
{\cal L}_g = g\; \epsilon^{(\lambda)}_\mu\cdot k^\mu\;.
\label{Lg}
\end{equation}
Vertex (\ref{Lg}) results in the imaginary part of the $f_{BD}$ constant,
so that $\Im m f_{BD}(q^2)\to 0$ at $q^2\to s_i$, and, hence,
$\Im m f_{BD}\ll \Re e f_{BD}$. Using the vector dominance, one can easily 
get the relation between $\Im m f_{BD}$ and the transversal correlator
determined by the $\epsilon^{(\lambda)}_\mu$ current of decay and the
mesonic current of $J_\nu$ \cite{10}
$$
\Im m \Pi^{tr}_{Fg}(q^2) = - \frac{M}{2}\; \Im m f_{BD}\;,
$$
where $\Im m \Pi^{tr}_{Fg}$ coincides the expression in (\ref{FF}) with the
substitution $F^2\to Fg$. Then the dispersion relation for the $f_{BD}$
function at $q^2=s_i=(m_B+m_D)^2$ gives
\begin{equation}
f_{BD} = \frac{1}{16\pi^2}\; \frac{Fg}{9}\biggl(\frac{4\mu_{BD}}{M}\biggr)^3
\; M v^3_{th}\;,
\label{F2}
\end{equation}
Comparing (\ref{F1}) with (\ref{F2}), one finds
\begin{equation}
g = \biggl(\frac{M}{4\mu_{BD}}\biggr)\; 12\pi \sqrt{\frac{3}{10v_{th}}}.
\label{F3}
\end{equation}

\section{Scaling relation and numerical estimates}
As has been mentioned, the $v_{th}$ value is determined by the threshold 
of production of the vector excitations for the heavy mesons, $B^{*+}$ and
$D^{*0}$, so
$$
v^2_{th}\approx \frac{1}{2\mu_{BD}}\; (\Delta m_B + \Delta m_D)\;,
$$ 
where $\Delta m_B =m_{B^*}-m_B,\; \Delta m_D= m_{D^*}-m_D$. In the
Heavy Quark Effective Theory (see review in \cite{11}), one has
$$
m_B\Delta m_B = m_D\Delta m_D= const.\;,
$$ 
independently of the heavy quark flavor with the accuracy up to 
corrections over $\Lambda_{QCD}/m_{B,D}$. Hence, one gets
\begin{equation}
v_{th}\cdot \mu_{BD} = const.
\label{Fv}
\end{equation}
Using (\ref{Fv}) and (\ref{F3}), one can easily obtain the scaling
relation for the decay constant of the heavy vector quarkonium with the mass
$m_B+m_D < M < m_{B^*}+m_{D^*}$
\begin{equation}
\frac{g^2}{M}\; \biggl(\frac{4\mu_{BD}}{M}\biggr) = const.
\label{SRel}
\end{equation}
Relation (\ref{SRel}) is in a good agreement with the experimenal data on
the ratio of constants for the decays of $\Upsilon(4S)\to B^+B^-$
and $\psi(3770)\to D^+D^-$, where one has the accuracy of $\Delta g \simeq 3$
(see table I). Note, that the estimate due to (\ref{F3}) giving
$g_{\Upsilon B\bar B}=57$ agrees the experimental value taken as the input 
parameter for the scaling relation. The latter fact points out the
self-consistency of the method resulting in (\ref{SRel}).
As for the accuracy of the scaling relation, it is determined by the 
uncertainty in the sum rules, where eq.(\ref{F3}) has been derived. Remember,
that the stability of the $f$ constant calculation over the initial 5 moments
of the spectral density changes from 5\% for $\Upsilon(4S)$ to 25\%
for $\psi(3770)$ with the decrease of the vector state mass. This must be 
included in the systematic uncertainty of the method used. We evaluate
$\Delta g/g \sim 15-20\%$ for $B_c^{*+}(3S)$, so that
$$
g_{B_cBD} = 49\pm 8\;.
$$
The decay width is determined by the expression
\begin{equation}
\Gamma(B_c^{*+}(3S)\to B^+D^0) = \frac{1}{24\pi}\; g^2\;
\frac{|{\bf k}|^3}{M^2} \approx 90\pm 35\;\; {\rm MeV.}
\label{FN}
\end{equation}
We assume that the channel of decay to $B^*D$ can be neglected, since
it is suppressed by the third power of the momentum of the decay final states
due to the greater mass of $B^*$ in comparison with the $B$ mass.
Then taking into account the channel $B^0 D^+$, the total width of 
$B_c^{*+}(3S)$ is equal to $\Gamma_{tot}=180\pm 70$ MeV. We have supposed 
$M(B_c^{*+}(3S))=7.250$ GeV \cite{5} in the numerical estimate of (\ref{FN}).
Note, that the width strongly depends on the difference of masses,
$\Delta M = M - (m_B+m_D)$ determining $|{\bf k}|$. At the used value of the
quarkonium mass, one has $\Delta M \sim 110$ MeV, which differs from
$\Delta M \sim 30$ MeV for the decays of $\Upsilon(4S)\to B^+B^-$
and $\psi(3770)\to D^+D^-$. The larger phase space results in the fact that
the total $B_c^{*+}(3S)$ width is one order of magnitude greater than the 
total widths of $\Upsilon(4S)$ and $\psi(3770)$ having $\Gamma_{tot}\simeq
24$ MeV.

\section{Conclusion}
In this paper we have considered the sum rules for the mesonic currents. These
sum rules allow one to determine the coupling constant of the heavy
vector $(\bar b c)$-quarkonium decaying to the heavy meson pair,
$$
g = \biggl(\frac{M}{4\mu_{BD}}\biggr)\; 12\pi \sqrt{\frac{3}{10v_{th}}}\;,
$$
where $m_B+m_D < M < m_{B^*}+m_{D^*}$. The value of $v_{th}$ determining
the threshold of the nonresonant contribution into the transversal 
correlator of currents, is given by the mass splitting between the vector 
and pseudoscalar states of heavy mesons, and it possesses the definite
scaling property, so that one has derived the relation
$$
\frac{g^2}{M}\; \biggl(\frac{4\mu_{BD}}{M}\biggr) = const.\;,
$$
which is in a good agreement with the experimental data on the constants
of decays of $\Upsilon(4S)\to B^+B^-$ and $\psi(3770)\to D^+D^-$.
The numerical estimate of the $B_c^{*+}(3S)\to B^+D^0$ decay width
strongly depends on the mass difference $\Delta M = M - (m_B+m_D)$ 
determining the phase space, so that at $M(B_c^{*+}(3S))=7.250$ GeV
one has found $\Gamma= 90\pm 35$ MeV.
\section*{Acknowledgements}
This work is in part supported by the Russian Foundation of Fundamental 
Researches, grant 96-02-18216, and by the Russian State Stipends for
young scientists. The author thanks academician S.S.~Gershtein for
the stimulating discussions and support.

\newpage
\begin{table}[p]
\caption{The predictions of scaling relation in comparison with the 
current experimental data}
\begin{tabular}{lcc}
value & exp. & scaling rel.\\
\hline
$g_{\Upsilon(4S)\to B^+B^-}$ & 52 & input\\
$g_{\psi(3770)\to D^+D^-}$   & 31 & 31\\
$g_{B_c^{*+}(3S)\to B^+D^0}$ & -- & 49\\
\end{tabular}
\end{table}

\end{document}